\documentclass[12pt,preprint]{aastex}

\shorttitle{Short Title} \shortauthors{Wang \& Fan}

\begin{document}


\title{The distribution of two-dimensional eccentricity of Sunyaev-Zeldovich
Effect and X-ray surface brightness profiles}


\author{Y.-G. Wang and Z.-H. Fan}
\affil{Department of Astronomy, Peking University,
    Beijing 100871, China}
\email{wangyg@bac.pku.edu.cn,fan@bac.pku.edu.cn}




\begin{abstract}
With the triaxial density profile of dark matter halos and the corresponding equilibrium
gas distribution, we derive two-dimensional Sunyaev-Zeldovich (SZ) effect and X-ray
surface brightness profiles for clusters of galaxies. It is found that the contour map
of these observables can be well approximated by a series of concentric ellipses
with scale-dependent eccentricities. The statistical distribution of their
eccentricities (or equivalently axial ratios) is analyzed by taking
into account the orientation of clusters with
respect to the line of sight and the distribution of the axial ratios and the concentration
parameters of dark matter halos. For clusters of mass
$10^{13}h^{-1}{ M}_{\odot}$
at redshift $z=0$, the axial ratio is peaked at $\eta \sim 0.9$ for both SZ and X-ray profiles.
For larger clusters, the deviation from circular distributions is more apparent,
with $\eta$ peaked at $\eta \sim 0.85$ for $M=10^{15}h^{-1}{ M}_{\odot}$.
To be more close to observations, we further study the axial-ratio distribution
for mass-limited cluster samples with the number distribution of clusters at different
redshifts described by a modified Press-Schechter model. For a mass limit of value
$M_{lim}=10^{14}h^{-1}{ M}_{\odot}$, the average axial ratio is
$\langle \eta \rangle \sim 0.84$ with a tail extended to $\eta \sim 0.6$.
With fast advance of high quality imaging observations
of both SZ effect and X-ray emissions, our analyses provide a
useful way to probe
cluster halo profiles and therefore to test theoretical halo-formation models.
\end{abstract}




%
\section{Introduction}
Following abundant observations of various kinds, a concordance
cosmological model has emerged (e.g., Spergel et al. 2003).
The geometry of the universe is nearly flat with the total
density of the universe $\Omega_{tot}$ very close to unity.
Among which, about $70\%$ is in the form of dark energy with
negative pressure, and about $25\%$ is composed of non-baryonic
dark matter. The familiar baryonic matter contributes only
about $5\%$ of the total matter density in the universe.
It is apparent that dark matter plays dominant roles in the
structure formation.

Besides large-scale structures, it is of great cosmological
interest to understand the properties of individual dark matter
halos. Extensive studies with high resolution simulations have
been performed in this aspect (e.g., Fukushige \& Makino 1997,
2001; Moore et al. 1998; Jing 2000; Jing \& Suto 2000). Navarro,
Frenk and White (1996, 1997) put forward a universal density
profile (NFW profile) to describe the spherical mass distribution
of dark matter halos of different masses. Confronted with
observations, the validity of the profile has been tested for
dwarf galaxies, galaxies, as well as clusters of galaxies. Certain
degree of disagreement between the inferred inner profile from
some observations and the NFW one raised a potential small-scale
problem for cold dark matter models. Despite that, the NFW profile
has been applied widely in different studies (e.g., Bartelmann et
al. 2001; Li \& Ostriker 2002; Wyithe et al. 2001). However, the
sphericity assumed in the NFW profile is clearly an approximation,
and could introduce systematic errors in theoretical predictions,
and further in cosmological parameter extractions from
observations (e.g., Birkinshaw 1999). Among others (Dubinski 1994;
Jing, et al. 1995; Thomas et al. 1998; Yoshida et al. 2000;
Meneghetti et al. 2001), Jing and Suto (2002) did detailed
analysis with numerical simulations, and proposed a NFW-like
triaxial density profile for dark matter halos with quantitative
statistics on the distribution of the axial ratios and the
concentration parameter. Their studies allow one to investigate
the systematic differences caused by the non-sphericity of mass
distributions. Significant effects have been found in the
statistics of arcs, and the probability of large separation
gravitational lenses (e.g., Oguri, Lee \& Suto 2003; Meneghetti,
Bartelmann \& Moscardini 2003; Dalal et al. 2004; Oguri \& Keeton
2004).

For clusters of galaxies, the distribution of intra-cluster gas
(ICM) is closely associated with the underlying dark matter
distribution, and therefore observations, such as X-ray (e.g.,
Rosati et al. 2002) and Sunyaev-Zeldovich (SZ) effect (e.g.,
Carlstrom et al. 2002) can provide us information on the dark
matter profiles (e.g., Allen et al. 2002; Grego et al. 2004; Grego
et al. 2001) . Lee \& Suto (2003) studied the equilibrium gas
distribution in triaxial dark matter halos under the approximation
of small ellipticities. They (Lee \& Suto 2004) further analyzed
the X-ray and SZ effect distributions associated with the triaxial
mass distribution of clusters of galaxies. Since they aimed at
finding the dark matter distribution for individual clusters, they
parametrized the observables (X-ray and SZ) with the parameters
being functions of those characterizing the mass profile of dark
matter halos. While this type of reconstruction is certainly
useful, individuality of clusters affects the results
considerably. On the other hand, for a large number of clusters,
one can study collectively the profile of the observables, and
further investigate the underlying dark matter distribution
statistically. Mohr et al. (1995) analyzed 65 X-ray clusters, and
presented statistical results on several quantities representing
the morphology of clusters' X-ray emissions. With their limited
sample, they found that the distribution of the axial ratio of
X-ray emissions has a main peak at $\eta \sim 0.9$ with a minor
peak at $\eta \sim 0.6$. The average axial ratio is $\langle
\eta\rangle \sim 0.8$. To investigate the three-dimensional ICM
distribution, Cooray (2000) assumed a spheroidal ICM profile
(prolate or oblate), and obtained statistics of 3-D axial ratio
given the 2-D results of Mohr et al. (1995). Similar studies on
elliptical galaxies have been carried out for many years (e.g.,
Binney 1985; Binney \& de Vaucouleurs 1981).

The triaxial mass distribution of dark matter obtained from
numerical simulations (Jing and Suto 2002) provides us a starting point
to study the ICM distribution without artificial assumptions. In this
paper, we analyze the corresponding surface profiles of X-ray
and SZ effects. Instead of focusing on individual clusters, we
investigate the profiles of observables statistically. Specifically
the axial-ratio distributions of X-ray and SZ effects
profiles resulting from the triaxial dark matter distributions are
derived. Our study presents another view to test the theory of the formation
of dark matter halos.

The rest of the paper is organized as follows. Under the equilibrium
assumption, in \S2, we derive the two-dimensional profiles of SZ and X-ray
observables from triaxial dark matter distributions. In \S3,
we obtain the two-dimensional eccentricities of the SZ and X-ray
profiles under the linear approximation for the deviation of sphericity.
Section 4 presents the statistical
analysis on the distribution of two-dimensional eccentricities,
which can be compared with observational results. Summary and discussion
are given in \S5. Appendix A contains derivations of the eccentricities
for the isothermal SZ effects. Results for other cases are also presented.
Appendix B discusses possible values of the parameter $C/K$ associated
with X-rays and SZ effects.

Throughout the paper, we consider the concordance cosmological model
with $\Omega_m=0.3$, $\Omega_{\Lambda}=0.7$, $H_0=72km/s/Mpc$ and
$\sigma_8=0.9$, where $\Omega_m$ and $\Omega_{\Lambda}$ are the
present dimensionless matter density and the dark energy density
of the universe, respectively,
$H_0$ is the present Hubble constant, and $\sigma_8$ is rms of the extrapolated
linear mass density fluctuation smoothed over $8h^{-1}Mpc$ with $h$ the
Hubble constant in units of $100km/s/Mpc$.

\section{Two-dimensional SZ and X-ray profiles}
\subsection{Gas density distribution}

The three-dimensional triaxial density distribution of dark matter halos
is given by Jing \& Suto (2002), which has the following form
\begin{equation}\label{w1}
\frac{\rho(R)}{\rho_{crit}}=\frac{\delta_c}{(R/R_0)^\alpha(1+R/R_0)^{3-\alpha}}
\end{equation}
where $R=a({x^2}/{a^2}+{y^2}/{b^2}+{z^2}/{c^2})^{1/2}\  (c\leq
b\leq a)$ is the length of the major axis, $R_0$
is a scale radius, $\delta_c$ is a characteristic density and $\rho_{crit}$
is the critical density of the universe. For the
value of $\alpha$, it is found that both $\alpha=1$ and $\alpha=1.5$
can fit well
to the profiles of simulated dark matter halos. Detailed comparison
showed that $\alpha=1$ is slightly better for cluster-scale halos while
for galactic halos, $\alpha=1.5$ gives better results.
In this paper, we focus on clusters of galaxies, and adopt $\alpha=1.0$.

Under the equilibrium assumption, we can derive the gas distribution
in a triaxial dark matter halo. The related equations are the
Poisson equation and the equilibrium equation. Specifically,
\begin{equation}
\bigtriangledown^2\Phi=4\pi G\rho \  ,
\end{equation}
\begin{equation}\label{w4}
\frac{1}{\rho_g}\bigtriangledown P_g=-\bigtriangledown \Phi \ ,
\end{equation}
where $\Phi$ is the gravitational potential, $\rho$ is the total matter
density, $\rho_g$ is the gas density, and $P_g$ is the gas pressure.

We consider both isothermal and polytropic ICM states. For the
isothermal case, we have
\begin{equation}\label{w5}
P_g=\frac{k_BT_g}{\mu m_p}\rho_g \equiv K\rho_g,
\,\,\,\,\,\,\,\,\,\,\,\,\,\,\,\,
\end{equation}
where $k_B$, $T_g$, $\mu$, and $ m_p$ represent the Boltzmann
constant, the gas temperature, the mean molecular weight, and the
proton mass, respectively. Then the gas density is related to the
gravitational potential through the following equation
\begin{equation}\label{w6}
\frac{{\rho}_{g}}{{\rho}_{g0}}={\exp{[-{\frac{1}{K}}(\Phi-{\Phi}_0)}]},
\end{equation}
where the subscript  $¡°0¡±$  denotes the corresponding value at the
central position.

For polytropic gas, the equation of state is
\begin{equation}\label{w7}
P_g=P_{g0}[\rho_g/\rho_{g0}]^{\gamma},
\end{equation}
where $\gamma$ is the polytropic index.
The equilibrium gas density can be derived as
\begin{equation}\label{w8}
{\rho}_g=\rho_{g0}\bigg \{ 1-\frac{1}{K_0}\frac{\gamma-1}{\gamma}
[{\Phi}-{\Phi}_0]\bigg \}^{\frac{1}{\gamma-1}},
\end{equation}
where $K_0=k_BT_{g0}/\mu m_p$ with $T_{g0}$ the central temperature
of ICM.
It is seen that the gas density profile is closely associated
with that of the gravitational potential, and the isopotential contours
are also isodensity contours of ICM.

As ICM contributes about $10\%$ of total cluster mass, it is neglected
in calculating the gravitational potential. From the triaxial
density profile of dark matter halos (eq.[1]), one can solve
the Poisson equation to obtain the gravitational potential numerically
(e.g., Binney \& Tremaine 1987).

Under the approximation of small eccentricities for dark matter halos,
Lee and Suto (2003) derived an analytical solution for the
gravitational potential, which is given as
\begin{equation}\label{w2}
\Phi({\vec u}) \approx
C{[F_1(u)+{\frac{{e_b}^2+{e_c}^2}{2}}F_2(u)+{\frac{{e_b}^2{\sin}^2{\theta}_0{\sin}^2{\varphi}_0+{e_c}^2{\cos}^2{\theta}_0}{2}}F_3(u)]},
\end{equation}
where $\vec u\equiv \vec r/R_0$, $ \vec r=(x,y,z)=r(\sin{\theta}_0\cos{\varphi_0},\sin{\theta}_0\sin{\varphi_0},\cos{\theta_0})$,
$C=4\pi G\delta_c\rho_{crit}{R_0}^2$,
$e_b=(1-{b^2}/{a^2})^{1/2}$ and $e_c=(1-{c^2}/{a^2})^{1/2}$.
Here $e_b$ and $e_c$ are the two eccentricities of
the ellipsoidal dark matter halos.
The three functions $F_1(u), F_2(u)$ and
$F_3(u)$ are given in Lee and Suto (2003). In the following, our analytical
analyses on the two-dimensional SZ and X-ray profiles are
obtained primarily from the potential given by equation (8).
On the other hand,
we also solve for the gravitational potential numerically from Poisson
equation, and derive the exact SZ and X-ray profiles. These
exact solutions are compared with those analytical ones to test the validity
of the approximation underlying equation (8).

For triaxial clusters, their appearances depend on the viewing
direction. With $(x,y,z)$ the principal coordinate system, we introduce
$(x^{\prime},y^{\prime},z^{\prime})$ with $z^{\prime}$ the line
of sight direction and $x^\prime$ lying in the $x-y$ plane (e.g., Binney 1985;
Oguri, Lee \& Suto 2003). The two systems are related by
\begin{eqnarray}\label{wf9}
\left[\matrix{
x\cr
 y\cr
  z
  }\right]
   =T\left[\matrix{
    x^{\prime} \cr
          y^{\prime} \cr
            z^{\prime}
        } \right],
        \end{eqnarray}
where
\begin{eqnarray}\label{w10}
 T\equiv \left[\matrix{
     -\sin{\phi} & -\cos{\phi}\cos{\theta} & \cos{\phi}\sin{\theta} \cr
         \cos{\phi} & -\sin{\phi}\cos{\theta} & \sin{\phi}\sin{\theta} \cr
         0            & \sin{\theta}             & \cos{\theta}
         } \right].
\end{eqnarray}
Here $\theta$ and $\phi$ are the polar coordinates of the line of sight
direction in the $(x,y,z)$ coordinate system. Then we have
\begin{eqnarray}\label{w11}
\Phi{(\vec u^{\prime})} &&\approx C \bigg [    F_1(u^{\prime})+{\frac{ {e_b}^2+{e_c}^2  } {2}  } { F_2(u^{\prime})} \nonumber \\
&& +{    \frac{  { {e_b}^2 {   (
x^{\prime}\cos{\phi}-y^{\prime}{\sin{\phi}\cos{\theta}+z^{\prime}\sin{\phi}\sin{\theta})
}^2    } + {e_c}^2{  (
y^{\prime}\sin{\theta}+z^{\prime}\cos{\theta} ) } ^2    } }{2{r^{\prime}}^2}}
{{F_3}(u^{\prime})}\bigg ],
\end{eqnarray}
where $\vec {u^{\prime}}=\vec {r^{\prime}}/R_0$.

\subsection{SZ and X-ray surface profiles}
The typical mass of a cluster of galaxies is
$10^{14}-10^{15} \hbox { M}_{\odot}$,
to which the gas component contributes about $10\%$.
Because of the deep potential well, the typical temperature of
ICM is a few $keV$. Therefore the gas is fully ionized. Hot electrons
emit X-ray primarily through thermal bremsstrahlung process.
On the other hand,
as cosmic microwave background (CMB) photons pass through
a cluster, they scatter with free electrons inside. As the result,
the CMB spectrum is distorted. This is the thermal SZ effect
(Sunyaev \& Zel'dovich 1970, 1972). Observing clusters
through X-ray and SZ effect has been getting mature, and
high resolution imaging of clusters of galaxies is possible.
Thus it is becoming observationally feasible to study
the ICM distribution and further the underlying dark matter distribution
in detail.

Here we derive the respective profiles of SZ effect and X-ray emission
from the gas distribution in a triaxial cluster.

The CMB spectrum distortion due to the thermal SZ effect can be
described by a frequency-dependent temperature change
$\delta T$ relative to the average CMB temperature.
In the nonrelativistic limit, the gas properties
affect the SZ effect through the Compton $y$-parameter, which is
proportional to the gas pressure along the line of sight. Specifically,
we have
\begin{equation}\label{w12}
\delta{T} \propto \int\! n_e{T_e} {\mathrm d}l,
\end{equation}
where $n_e$ and $T_e$ are the number density and temperature of electrons,
respectively. The integration is along the line of
sight.

For fully ionized gas primarily consisting of hydrogen and
helium, $n_e\propto \rho_g$. Thus
\begin{equation}\label{w12}
\delta{T} \propto \int\! \rho_g{T_e} {\mathrm d}l.
\end{equation}

For isothermal ICM, the SZ effect can then be written as
\begin{eqnarray}\label{w13}
\delta T(x^{\prime},y^{\prime}) &\propto &\int\!\rho_gT_e{\mathrm d}l
 \propto \int\!  {\exp}\bigg [-{\frac{1} {K}}(\Phi-\Phi_0)\bigg ] {\mathrm d}l
 \propto \int\! {\exp}\bigg [-\frac{1}{ K} \Phi\bigg ]{\mathrm d}l \nonumber\\
& \propto & \int\! {\exp}{\bigg \{}-\frac{C}{K}\bigg [F_1(u^{\prime})+\frac{{e_b}^2+{e_c}^2}{2}F_2(u^{\prime})   \nonumber\\
& &  +{    \frac{  { {e_b}^2 {   ( x^{\prime}\cos{\phi}-y^{\prime}{\sin{\phi}\cos{\theta}+z^{\prime}\sin{\phi}\sin{\theta}) }^2    } + {e_c}^2{  ( y^{\prime}\sin{\theta}+z^{\prime}\cos{\theta} ) } ^2    } }{2{r^{\prime}}^2}}\nonumber\\
&&  {{F_3}(u^{\prime})}   \bigg ] {\bigg \}}{\mathrm d}{z^ \prime}.
\end{eqnarray}

For polytropic gas, the SZ effect depends on the potential through
the following equation
\begin{equation}\label{w17}
\delta{T}(x^{\prime},y^{\prime})\propto \int\!\bigg \{1-\frac{1}{K_0}\frac{\gamma-1}{\gamma}
[{\Phi}(x^{\prime},y^{\prime},z^{\prime})-{{\Phi}}_0]\bigg \}^{\frac{\gamma}{\gamma-1}}{\mathrm
d}z^{\prime}.
\end{equation}

For bremsstrahlung X-ray emission, its surface brightness has a different
dependence on the density and the temperature of electrons, which is
\begin{equation}\label{w14}
S_x\propto\int{n_e}^2\Lambda_{eH}{\mathrm d}l,
\end{equation}
where $\Lambda_{eH}$ is the X-ray cooling function. For
bolometric bremsstrahlung emission, we adopt an
approximate expression $\Lambda_{eH}\propto{T_e}^{1/2}$.
In the case of isothermal gas, the above equation is reduced to
\begin{equation}\label{w15}
S_x\propto\int{n_e}^2{\mathrm d}l,
\end{equation}
and then we get
\begin{eqnarray}\label{w16}
S_X(x^{\prime},y^{\prime}) &\propto & \int\! {\exp}{\bigg \{}-\frac{2C}{K}\bigg [F_1(u^{\prime})+\frac{{e_b}^2+{e_c}^2}{2}F_2(u^{\prime})  \nonumber\\
& &  +{    \frac{  { {e_b}^2 {   ( x^{\prime}\cos{\phi}-y^{\prime}{\sin{\phi}\cos{\theta}+z^{\prime}\sin{\phi}\sin{\theta}) }^2    } + {e_c}^2{  ( y^{\prime}\sin{\theta}+z^{\prime}\cos{\theta} ) } ^2    } }{2{r^{\prime}}^2}}\nonumber\\
&&  {{F_3}(u^{\prime})} \bigg ] {\bigg \}}{\mathrm d}{z^ \prime}.
\end{eqnarray}

For polytropic gas, we have
\begin{equation}\label{w18}
S_X(x^{\prime},y^{\prime})\propto\int\!\bigg \{1-\frac{1}{K_0}\frac{\gamma-1}{\gamma}[{\Phi}(x^{\prime},y^{\prime},z^{\prime})-{{\Phi}}_0]\bigg \}^{\frac{3+\gamma}{2(\gamma-1)}}{\mathrm
d}z^{\prime}.
\end{equation}

In next section, we demonstrate that both $\delta T(x^{\prime},y^{\prime})$ and $S_X(x^{\prime},y^{\prime})$ can be well described
by a series of concentric ellipses, and we further derive an approximate
analytic expression for their axial ratios.

\section{Two-dimensional eccentricities of SZ effect and X-ray surface
brightness profiles}

Lee and Suto (2003) showed that the 3-D gas distribution in
a triaxial dark matter halo is approximately ellipsoidal with
decreasing eccentricities from the center to the outer part of the
cluster. Therefore we expect elliptical-like profiles for
both SZ effects and X-ray emissions.

Let $\xi$ be defined as:
\begin{eqnarray}\label{w19}
{\xi}^2
   = {\frac{1}{{R_{0}}^2}}
(\frac{{x^{\prime\prime}}^2}{1+\Delta}+\frac{{y^{\prime\prime}}^2}{   1+
{\Delta}^{\prime}}),
    \end{eqnarray}
where the coordinates $(x^{\prime\prime},y^{\prime\prime})$ relate to
$(x^{\prime},y^{\prime})$ by
\begin{eqnarray}\label{w20}
\left[\matrix{
    x^{\prime} \cr
        y^{\prime} \cr
     } \right]
         = \left[\matrix{
             \cos{\Psi} & -\sin{\Psi}  \cr
             \sin{\Psi} & \cos{\Psi}  \cr
             }  \right]
                   \left[\matrix{
                     x^{\prime\prime} \cr
                       y^{\prime\prime} \cr
                       }  \right].
                       \end{eqnarray}

The form $\xi$ given in equation (20) represents
that the length of both the major and minor axes is different from
the circular radius
$\tilde r^{\prime \prime}=\sqrt{x^{\prime\prime 2}+y^{\prime\prime 2}}$.
>From the view point of small aspherical perturbations,
a circle of $\tilde r^{\prime \prime}$ is perturbed
to an ellipse labeled by $\xi$. On the other hand,
for an elliptical form
$\tilde \xi^2=(1/R_0^2)[x^{\prime \prime 2}+y^{\prime \prime 2}/
(1+\Delta^{\prime \prime})]$,  only
one of the two axes is perturbed away from the circle.
In other words, with same $\xi$ and $\tilde \xi$, the corresponding
original circles $\tilde r^{\prime \prime}$ are different.
>From Appendix A (A13), we see that $\xi$ in equation (20) is in
general consistent with the perturbative results. Only
in the case of $e_c=0$ or $\theta=0$, $\tilde \xi$
gives a valid description.

To approximate the distributions of SZ effects or X-ray emissions
with a series of ellipses, we assume that the 2-D observables
are functions of $\xi$ only. Thus $\Psi$ represents
the orientation of an elliptical contour in $(x^{\prime},y^{\prime})$
coordinate system. With tedious but straightforward derivations,
which are shown in Appendix A for the isothermal SZ effects,
we obtain, to the leading order of $e_b^2$ and $e_c^2$,
the following general expressions for all the cases
considered
\begin{equation}\label{w36}
\tan {2\Psi}=\frac{C_{13}}{C_{11}-C_{12}},
\end{equation}

\begin{equation}\label{w37}
\Delta=-\bigg [-C_{13}\sqrt{1+\bigg (\frac{C_{11}-C_{12}}{C_{13}}\bigg )^2}+(C_{11}+C_{12}) \bigg ]M( \tilde u^\prime),
\end{equation}

\begin{equation}\label{w38}
\Delta^\prime=-\bigg [C_{13}\sqrt{1+\bigg (\frac{C_{11}-C_{12}}{C_{13}}\bigg )^2}+(C_{11}+C_{12})\bigg ] M(\tilde u^\prime),
\end{equation}
and
\begin{eqnarray} \label{w39}
{E}^2=1-\frac{1+\Delta}{1+{\Delta}^\prime}
=\frac{{\Delta}^\prime-\Delta}{1+{\Delta}^\prime}
\approx{\Delta}^\prime-\Delta=2C_{13}\sqrt{
 1+\bigg (\frac
  {C_{11}-C_{12}}{C_{13}}\bigg )^2 } [-M(\tilde u^\prime)],
   \end{eqnarray}
where $E$ is the eccentricity, and
\begin{equation}\label{w29}
C_{11} = -{e_b}^2\cos^2\phi,
\end{equation}
\begin{equation}\label{w30}
C_{12} = -({e_b}^2\sin^2\phi\cos^2\theta+{e_c}^2\sin^2\theta),
\end{equation}
\begin{equation}\label{w31}
C_{13} ={e_b}^2\sin{2\phi}\cos\theta,
\end{equation}
and $M$ is a function of $\tilde u^\prime=(\sqrt{x^{\prime 2}+y^{\prime 2}})/R_0$
only. The specific functional forms of $M$ for different cases can be found
in Appendix A.

>From the above perturbative results, we see explicitly
that for a relaxed cluster of galaxies,
$\Psi$ is independent of $\tilde u^{\prime}$, i.e.,
the elliptical contours of an observable are
in alignment with each other. For the eccentricity, its
$\tilde u^{\prime}$ and line-of-sight $(\theta, \phi)$ dependences
are separable. The $\tilde u^{\prime}$-dependent factor $M$
reflects the differences
between the distribution of dark matter and that of
the ICM, and is specific for different
quantities, SZ effects or X-ray emissions.

It is clear to see that the 2-D eccentricity $E$ is on the order
of $e_b$ or $e_c$ of the underlying dark matter halo.
For a cluster, the value of $E$ is very
similar for the four considered cases.
Figure 1 shows the eccentricity of the isothermal SZ profiles
at different radius. The parameters of
the cluster are chosen to be $M=10^{14}h^{-1}{ M}_{\odot}$,
$e_b=0.6$, and $e_c=0.8$. The concentration parameter is taken to be the
corresponding average value (Jing \& Suto 2002).
The line-of-sight direction lies in $\theta=0$ and $\phi=0$.
For this set of parameters, the perturbative result is
$E=e_b\sqrt{[-2M(\tilde u^{\prime})]}$, which is shown as the solid line.
The dotted line is the exact result from direct numerical calculations
without approximations. It is seen that the perturbative
eccentricity is systematically higher than that of the
numerical one, but not by a large fraction. The differences between
the two do not exceed $\sim 17\%$ for $\tilde u^\prime$ up to $10$.
In terms of the axial ratio $\eta$, the corresponding differences
are less than $\sim 5\%$. For smaller $e_b$ and $e_c$, the
perturbative and numerical results agree better.
Thus our approximate expressions describe the profiles of SZ effect
and X-ray emission reasonably well.
There is a mild change
of $E$ with radius. At the central region $E\sim 0.5$,
which corresponds to the axial ratio $\eta \sim 0.87$.
In the region around the virial radius, $E\sim 0.38$ and
$\eta \sim 0.93$.

Stark (1977) studied triaxial models of the bulge of M31.
Assuming an ellipsoidal volume brightness distribution with
constant axial ratios, $x^2+(uy)^2+(tz)^2=a_v^2$,
he derived the surface isophotes, and
obtained a series of similar ellipses with their axial ratio
depending on $u$, $t$ and the line-of-sight
direction. Specifically for the line-of-sight direction
shown in Figure 1, the axial ratio of isophotes is
$\beta=1/u$, or the eccentricity $\sqrt{1-(1/u)^2}$.
In the problem we analyze here, it is the dark matter halo
that has an ellipsoidal configuration with constant $e_b$ and
$e_c$. For the equilibrium ICM, it distributes more
spherically than that of the dark matter with scale-dependent
asphericity. Then our result $E=e_b\sqrt{[-2M(\tilde u^{\prime})]}$
can be clearly understood: $e_b$ is the eccentricity of
the dark matter halo in the line-of-sight direction, and
the factor $\sqrt{[-2M(\tilde u^{\prime})]}$ shows the differences
between the distributions of dark matter and ICM.
The value of $E$ is smaller than $e_b=0.6$.

\section{Statistical analysis for the eccentricity distribution}

In this section, we investigate the expected statistical distributions
of $E$ (or equivalently the axial ratio $\eta$) of 2-D observables
if dark matter halos of clusters of galaxies do have triaxial
mass distributions as revealed by numerical simulations.
Our theoretical results can be potentially confronted with observational
results, and the agreement/disagreement between the two can be used
as evidences supporting/challenging the cold dark matter structure
formation scenario.

As shown in equation (25),
the eccentricity is a function of $(e_b, e_c, c_e, \theta, \phi)$.
Thus its statistical distribution can be calculated through the
distributions of the five quantities. For a sample of clusters
of mass $M$ at redshift $z$, the differential distribution
function $p(E^2)$ has the form
\begin{equation}\label{w45}
p(E^2)dE^2=\bigg [\frac{2}{\pi}\int\!p(c/a){\mathrm d}(c/a)\int\!
p(c/b\mid c/a){\mathrm d}(c/b)\int\!p(c_e){\mathrm
d}c_e\int\!\bigg ({\partial{E^2}\over \partial {\phi}}\bigg )^{-1}_{\theta,c_e,c/b,c/a}
\sin{\theta}{\mathrm d}\theta \bigg ]dE^2,
\end{equation}
where the inner most integral reflects the projection effect, i.e.,
the change of the appearance of a fixed cluster due to different viewing.
The other three integrals are for the distributions of the concentration
parameter $c_e$, and the two axial ratios $c/a$ and $c/b$ given by
Jing and Suto (2002).

Before going to the full distribution of equation (29), we first
study the distribution of $E^2$ due solely to the
projection effect, which can be calculated through
\begin{eqnarray}\label{w40}
p(E^2\mid
c_e,c/b,c/a)&=&{2\over \pi}\int_{\theta_1}^{\theta_2}
\bigg (\frac{\partial E^2}{\partial\phi}\bigg )^{-1}_{\theta, c_e, c/b, c/a}\sin{\theta}{\mathrm
 d}\theta
 \end{eqnarray}
where $(\theta_1,\theta_2)$ is the range of $\theta$ in which
a physically meaningful value of $\phi$ can be found for
a fixed value of ${E}^2$ (Binney \& de Vaucouleurs 1981).

In Figure 2, we show $p({E}^2\mid c_e,c/b,c/a)$
of the isothermal SZ effect for a cluster of
mass $M=10^{14}h^{-1}M_{\odot}$ at $z=0$. Three sets of $(e_b,e_c)$
are considered. In all the cases, there are two peaks in the distribution,
which correspond to the two largest apparent axial ratios when viewed
along the cluster's principal axes. Specifically in our definition
of the coordinate systems, the two line of sights are $\theta=0$ and
($\theta=\pi/2, \phi=0$), respectively. Our results are consistent with
the study of Binney and de Vaucouleurs (1981) for elliptical
galaxies. Note that $e_b$ and $e_c$ are the eccentricities of the dark matter
halo and the 3-D gas profile is rounder than that of the
dark matter distribution.

Now let us consider the full probability function.
The relevant distributions are taken from Jing and Suto (2002).
Specifically we adopt (Oguri, Lee \& Suto 2003)
\begin{equation}\label{w41}
p(c/a)=\frac{1}{\sqrt{2\pi}\times0.113}\exp \bigg \{ -\frac{
[(c/a)(M_{vir}/M_\star)^{0.07\Omega(z)^{0.7}}-0.54]^2}{2(0.113)^2}\bigg \}\bigg (\frac{M_{vir}}{M_\star}\bigg)^{0.07{\Omega(z)}^{0.7}},
\end{equation}
where $M_\star$ is the characteristic nonlinear mass scale satisfying
$\sigma(M_\star)=1.68$ with $\sigma(M_\star)$ the rms of linear
density fluctuations smoothed over mass scale $M_\star$, and
\begin{equation}\label{w42}
p(c/b\mid
c/a)=\frac{3}{2(1-r_{min})}\bigg [1-\bigg (\frac{2c/b-1-r_{min}}{1-r_{min}}\bigg )^2\bigg],
\end{equation}
for $c/b\geq r_{min}$, where $r_{min}=c/a$ for $c/a\geq0.5$ and
$r_{min}=0.5$ for $c/a<0.5$. For $c/b\leq r_{min}$, $p(c/b\mid
c/a)=0$. For the concentration parameter $c_e$, we have
\begin{equation}\label{w43}
p(c_e)=\frac{1}{\sqrt{2\pi}\times0.3}\exp\bigg [-\frac{(\ln{c_e}-\ln{{\overline{c}}_e})^2}{2(0.3)^2}\bigg ]\frac{1}{c_e},
\end{equation}
where the fitting formula of ${\overline{c}}_e$ in the triaxial
model is given by:
\begin{equation}\label{w44}
{\overline{c}}_e=1.35\exp\bigg \{-\bigg [
\frac{0.3}{(c/a)(M_{vir}/M_{\star})^{0.07\Omega(z)^{0.7}}} \bigg ]^2\bigg \}
A_e\sqrt{\frac{\Delta_{vir}(z_c)}{\Delta_{vir}(z)}
\frac{\Omega(z)}{\Omega(z_c)}}\bigg (\frac{1+z_c}{1+z}\bigg)^{3/2}
\end{equation}
where $z_c$ is the collapse redshift of the cluster of mass
$M_{vir}$ and $\Delta_{vir}(z)$ is the average density of a virialized
halo with respect to the critical density, i.e.,
$\Delta_{vir}(z)\equiv (3M_{vir})/(4\pi r_{vir}^3\rho_{crit})$.
For the numerical factor $A_e$, we take $A_e=1.1$.

In Figure 3, the probability functions are shown for the
four cases we discussed. To be more comparable with the results of
Mohr et al. (1995), here and after we change $p(E^2)$
to $f(\eta)$, the probability function of the axial ratio $\eta$.
The cluster mass is taken to be $M_{vir}=10^{14}h^{-1}M_{\odot}$ and
is placed at redshift $z=0$. For all the four cases,
$f(\eta)$ are very similar with a peak at $\eta \sim 0.9$ and
a tail extended to $\eta \sim 0.6$. In the following study,
we concentrate on $f(\eta)$ of the isothermal SZ effect.

Figure 4 demonstrates the dependence of $f(\eta)$ on the cluster mass
(left) and on the redshift (right). For larger clusters, the SZ profiles
become more elongated, with the peak of $f(\eta)$ moving
from $\eta\sim 0.9$ for $M_{vir}=10^{14}h^{-1}M_{\odot}$ to
$\eta\sim 0.85$ for $M_{vir}=10^{15}h^{-1}M_{\odot}$.
On the other hand, $f(\eta)$ does not show significant redshift evolution.

The above statistics focused on cluster samples with fixed masses
and at fixed redshifts. Observationally, a cluster sample is
often flux limited so that the sample contains clusters of various masses
at different redshifts. Under the equilibrium assumption,
a flux limit (SZ effect or X-ray) can be converted to
a mass limit which is redshift dependent. Thus
in the following we consider mass limited cluster samples.
For simplicity, however, we choose a redshift-independent
mass limit $M_{lim}$ for each sample.
To certain extents, SZ-selected clusters are more or less
fixed-mass-selected clusters (Holder et al. 2000).
For a mass limited sample, the probability density $f(\eta)$ can be
calculated through
\begin{equation}
f(\eta)={\int dV(z) \int_{M_{lim}} f(\eta)|_{M,z}(dn/ dM )dM
\over \int  (dn/ dM) dM dV},
\end{equation}
where $f(\eta)|_{M,z}$ is the probability density of $\eta$
for clusters of fixed mass $M$ at redshift $z$ analyzed
above, $(dn/dM)dM$ is the number density of clusters of mass $(M,M+dM)$,
and $dV$ is the volume element.

For the number density of clusters, we adopt the one from
Jenkins et al. (2001)
\begin{eqnarray}\label{w46}
 \frac{dn}{dM}(z,M)&=&0.315\frac{\rho_0}{M}\frac{1}{\sigma_M}\mid{\frac{d\sigma_M}{dM}\mid}\nonumber\\
 &&\times\exp[-\vert 0.61-\ln(D_z\sigma_M)\vert^{3.8}],
 \end{eqnarray}
where $\rho_0$ is the present matter density of the
universe, $D_z$ is the linear growth factor, and $\sigma_M$
is the rms of the linearly extrapolated-to-present matter density
fluctuation over the mass scale $M$, which is given by Viana \& Liddle
(1999)
 \begin{equation}\label{w47}
  \sigma_M=\sigma_8\bigg [\frac{R(M)}{8h^{-1}Mpc}\bigg ]^{-\gamma[R(M)]}
 \end{equation}
where $R(M)$ is the comoving radius corresponding to $M$, and
$\gamma$ has the following form

  \begin{equation}\label{w49}
   \gamma(R)=(0.3\Gamma+0.2)\bigg [2.92+\log_{10}
   \bigg (\frac{R}{8h^{-1}Mpc}\bigg )\bigg ]
   \end{equation}
with $\Gamma$ the shape parameter.

In the left panel of Figure 5, the mass limited
$f(\eta)$ of the isothermal SZ effect
is shown for $M_{lim}=10^{13}h^{-1}M_{\odot}$ and
$M_{lim}=10^{14}h^{-1}M_{\odot}$, respectively. The axial ratio is
calculated at the virial radius $r_{vir}$.
It is seen that for the cluster sample with a larger $M_{lim}$, the
distribution $f(\eta)$ has a peak at a smaller $\eta$ and extends more
toward elongated configurations. The average axial ratio $\langle \eta \rangle\approx
0.87$ and $0.84$ for $M_{lim}=10^{13}h^{-1}M_{\odot}$ and
$M_{lim}=10^{14}h^{-1}M_{\odot}$, respectively. Although only
the results for the SZ effect are shown here, we expect similar
$f(\eta)$ for X-ray emission profiles.

The measurements of Mohr et al.(1995) on the emission-weighted axial
ratio for $57$ X-ray clusters gave $\langle \eta \rangle\approx 0.8$.
The distribution has a major peak at $\eta \sim 0.9$ and a minor
one at $\eta \sim 0.6$. They compared the
results with those of numerical simulations of
different cosmologies. Due to the strong dynamical evolution of clusters,
the profiles of X-ray emissions show significant non-sphericity
for $\Omega_m=1$ model with $\langle \eta \rangle\approx 0.7$.
For low density models, the late time evolution of clusters is weak,
and the clusters are nearly spherical with $\langle \eta \rangle\approx 0.95$ for
$(\Omega_m=0.2, \Omega_{\Lambda}=0)$ and $\langle \eta \rangle\approx 0.91$ for
$(\Omega_m=0.2, \Omega_{\Lambda}=0.8)$.
The observational $f(\eta)$ lies in between the models of
$\Omega_m=1$ and $\Omega_m=0.2$. Our analysis presented here is for
the cosmological model $(\Omega_m=0.3, \Omega_{\Lambda}=0.7)$.
For $M_{lim}=10^{14}h^{-1}M_{\odot}$, we have $\langle \eta \rangle\approx 0.84$,
which is very close to the observational value of $\langle \eta \rangle\approx 0.8$.
On the other hand, there are several differences between
our idealized cluster sample and the observational sample of
Mohr et al. (1995), so that the comparison of the two is tentative.
Firstly, their sample is more or less
a flux limited sample with some complications (Mohr et al. 1995;
Edge et al. 1990), while our analysis is performed on a
mass limited sample. Secondly, their statistics is on
the emission-weighted axial ratio that emphasizes more on the
core structure. Our results in the left panel of
Figure 5 are for $\tilde r^{\prime}=r_{vir}$.
We have known that the ellipticities of the SZ effect and
X-ray profiles are higher for the central region of a cluster
than that of the outer part (see Figure 1). Therefore we expect
smaller $\langle \eta \rangle $ for the emission-weighted axial
ratio in our model, which would be closer to the observational results.
For the purpose of demonstration, in the right panel of
Figure 5, we show the results
for $\tilde r^{\prime}=r_{vir}$ and $\tilde r^{\prime}=0.7r_{vir}$,
respectively, for
$M_{lim}=10^{13}h^{-1}M_{\odot}$. The respective average axial ratios are
$\langle \eta \rangle \approx 0.87$ and $0.86$.

The minor peak revealed by Mohr et al. (1995) does not show up
in our theoretical analysis. Whether this is a serious problem
for the model is not clear since the number of observed clusters ($57$)
is still not large enough for a solid statistical conclusion on the
appearance of the low peak. This minor peak may just represent
the tail of the distribution.

\section{Summary and Discussion}

X-ray emission and SZ effect from a cluster of galaxies
are directly related to the hot ICM, which
is gravitationally bound to the deep potential well of dark matter
halos of clusters. Therefore, different from optically selected clusters whose
identifications are affected considerably by the projection effects,
X-ray or SZ selected clusters are relatively clean.
Further more, the continuous distribution makes the ICM
a better tracer of the overall underlying dark matter
than galaxies. Thus X-ray and SZ observations have been
used extensively to study the distribution of dark matter (e.g.,
Buote and Lewis 2004).

Based on the triaxial dark matter halo model from numerical
simulations, we derived the profiles of SZ effects and X-ray
emissions for clusters of galaxies.
It is found that they can be well described
by a set of concentric ellipses with the eccentricity $E$ decreasing
toward the outer part of a cluster. Under the approximation that
the triaxiality of dark matter halos is weak, we obtained
an analytical expression for eccentricities of these observables,
which is a function of $e_b$, $e_c$, $c_e$, $\theta$ and $\phi$.
Here the first three quantities represent the intrinsic shape
of dark matter halos, and the last two label the line-of-sight
direction. For the four cases we discussed (SZ/X-ray, isothermal/polytropic),
the elongation of the profiles is similar.
For a halo of $M=10^{14}h^{-1}M_{\odot}$
and $e_b=0.6$ and $e_c=0.8$, $E \sim 0.4$ for $\theta=0$ and
$\phi=0$.

Our analytical results of (22) to (25) can be potentially used
to constrain the dark matter distribution of a cluster from
the corresponding observations of $\Psi$ and $E$. Unfortunately,
the line-of-sight direction $(\theta, \phi)$ cannot be known
beforehand, which strongly limits the reconstruction of
dark matter distribution from X-ray and SZ observations for
individual clusters. However, given $\Psi$ and $E$,
certain constraints on $(\theta, \phi)$ and therefore $e_b$
and $e_c$ can be obtained. For example, as shown in Figure 1,
with $\Psi=0$, the observed $E$ at the central region of a cluster
puts a lower limit on $e_b$.

On the other hand, analyses on the distribution of $E$ for a sample of clusters
can probe the asphericity of dark matter distribution statistically.
With the knowledge of the distributions of the five
parameters, we calculate the statistics
of the 2-D eccentricity or equivalently the axial ratio $\eta$.
We found that for clusters of mass $M=10^{13}h^{-1}M_{\odot}$,
$\langle \eta \rangle \sim 0.89$, and $\langle \eta \rangle \sim 0.85$ for $M=10^{15}h^{-1}M_{\odot}$.
The distribution $f(\eta)$ depends very weakly on the redshift.
We further studied $f(\eta)$ for mass-limited cluster samples. For
$M_{lim}=10^{13}h^{-1}M_{\odot}$, $\langle \eta \rangle \sim 0.87$. For
$M_{lim}=10^{14}h^{-1}M_{\odot}$, $\langle \eta \rangle \sim 0.84$.

Keeping in mind many differences between our idealized cluster samples
and the real observational one, we made a tentative comparison
with the emission-weighted $\eta$ distribution of Mohr et al. (1995)
for $57$ X-ray clusters. Their statistics showed that $\langle \eta \rangle\sim 0.8$. Our
analysis for $M_{lim}=10^{14}h^{-1}M_{\odot}$ gave $\langle \eta \rangle \sim 0.84$ at
$r=r_{vir}$. Since emission-weighted $\eta$ is influenced more
by the central part of a cluster, a smaller $\langle \eta \rangle $ is expected
if a similar weighting is done in our theoretical analysis, which
would bring a better agreement between the theoretical result and
the observational one.

Comparing with results of numerical simulations on different
cosmologies, Mohr et al. (1995) concluded that low density models
with $\Omega_m=0.2$ cannot explain the observed morphology of
clusters of galaxies. Although our theoretical analysis based
on the concordance model with $\Omega_m=0.3$ and $\Omega_{\Lambda}=0.7$
shows a certain degree of consistency with that of observations,
same discrepancies as noted by Mohr et al. (1995) are apparent.
The model tends to produce more spherical clusters than the observed ones.
Gravitationally it is related to the weak dynamical evolution of
dark matter halos for low-density models. In our study,
hydrostatic equilibrium is assumed for ICM, and thus the gas distribution
is determined fully by the dark matter distribution. In reality,
physical processes other than gravity can affect properties of ICM,
and change its equilibrium profile considerably.

On the other hand, the observational sample of Mohr et al. (1995)
contains $65$ clusters (the axial ratio statistics is presented
for $57$ clusters), and it is still not large enough for
a solid conclusion on the morphological statistics.
Currently, a large number of high resolution images of clusters
are available. Detailed studies on their morphologies are definitely
desirable. Because of the much better quality of images,
we expect that tighter constraints on theoretical models of structure
formation can be obtained through morphological studies on clusters of
galaxies.

\acknowledgments
We sincerely thank the referee for the constructive and detailed comments and
suggestions. This research was supported in part by the National Science Foundation of China
under grants 10243006 and 10373001, and by the Ministry of Science
and Technology of China under grant TG1999075401.

\appendix
\section{Appendix}
In this appendix, we present the derivation for
the 2-D eccentricity under the approximation of small $e_b$ and $e_c$.

To the linear order of $\Delta$ and $\Delta ^\prime $, from equation (20) we have
\begin{eqnarray}\label{w21}
\xi &\approx& \frac{1}{{R_0}}{({x^{\prime\prime}}^2-\Delta{x^{\prime\prime}}^2+
{y^{\prime\prime}}^2-{\Delta} ^{\prime}{y^{\prime\prime}}^2)}^{\frac{1}{2}}\nonumber\\
     &=&\frac{\sqrt{{x^{\prime\prime}}^2+{y^{\prime\prime}}^2}}{R_0}
     (1-\frac{\Delta{x^{\prime\prime}}^2+{\Delta}^{\prime}{y^{\prime\prime}}^2}
{{x^{\prime\prime}}^2+{y^{\prime\prime}}^2}
     )^{\frac{1}{2}}\nonumber\\
     &\approx&\frac{r^{\prime\prime}}{R_0}
     (1-\frac{1}{2}\frac{\Delta{x^{\prime\prime}}^2+{\Delta}^{\prime}{y^{\prime
\prime}}^2}{{{\tilde r}^{\prime\prime 2}}}),
\end{eqnarray}
where ${\tilde r^{\prime\prime 2}}={x^{\prime\prime}}^2+{y^{\prime\prime}}^2$.
In $(x^{\prime},y^{\prime})$, the above equation becomes
\begin{eqnarray}\label{w23}
\xi& =&\tilde u^{\prime}\bigg [1-\frac{\Delta(x^\prime\cos\Psi+y^\prime\sin\Psi)^2}{2{\tilde r^{\prime 2}}}-\frac{\Delta ^{\prime}(y^\prime\cos\Psi-x^\prime\sin\Psi)^2}{2{\tilde r^{\prime 2}}}\bigg ]\nonumber\\
&=&\tilde u^{\prime}\bigg [1-\frac{(\Delta \sin^2\Psi+\Delta ^{\prime}\cos^2\Psi)\sin^2{\theta ^\prime}+(\Delta \cos^2\Psi+\Delta  ^\prime \sin^2\Psi)\cos^2{\theta ^\prime}}{2}\nonumber\\
& &-\frac{(\frac{1}{2}\Delta\sin{2\Psi}-\frac{1}{2}\Delta^\prime\sin{2\Psi})\sin{2\theta ^\prime}}{2}\bigg ],
\end{eqnarray}
where ${\tilde u^{\prime}}={{\tilde r}^\prime}/{R_0}={\sqrt{{x^{\prime}}^2+{y^{\prime}}^2}}/{R_0}$, and
$(x^\prime,y^\prime)=(\tilde r^\prime\cos{\theta^\prime},\tilde r^\prime\sin{\theta^\prime})$
.

Our task is to determine the corresponding $\Delta$, $\Delta^{\prime}$
and $\Psi$ from the profiles derived from the triaxial dark matter
distribution. For all the four cases discussed above, the
derivations are similar.
In the following, we focus on the isothermal SZ profiles.

For $\delta T(x^{\prime},y^{\prime})=\delta T(\xi)$, we expand $\delta T(\xi)$
with respect to $\Delta$ and $\Delta^\prime$. To the linear oder, we have
\begin{eqnarray} \label{w24}
\delta T(\xi)& \approx  &  \delta T\bigg \{\tilde u^{\prime}\bigg [1\nonumber- \frac{(\Delta \sin^2\Psi+\Delta ^{\prime}\cos^2\Psi)\sin^2{\theta ^\prime}+(\Delta \cos^2\Psi+\Delta  ^\prime \sin^2\Psi)\cos^2{\theta ^\prime}}{2}\nonumber\\
& &-\frac{(\frac{1}{2}\Delta\sin{2\Psi}-\frac{1}{2}\Delta^\prime\sin{2\Psi})\sin{2\theta ^\prime}}{2}\bigg ]\bigg \}\nonumber\\
& \approx &\delta T(\tilde u^{\prime})-\frac{\partial\delta T(\tilde u^{\prime})}{\partial\tilde u^{\prime}}\tilde u^{\prime}\bigg [\frac{(\Delta \sin^2\Psi+\Delta ^{\prime}\cos^2\Psi)\sin^2{\theta ^\prime}+(\Delta \cos^2\Psi+\Delta  ^\prime \sin^2\Psi)\cos^2{\theta ^\prime}}{2}\nonumber\\
& &+\frac{(\frac{1}{2}\Delta\sin{2\Psi}-\frac{1}{2}\Delta^\prime\sin{2\Psi})\sin{2\theta ^\prime}}{2}\bigg]
\end{eqnarray}

On the other hand, with $f(u^\prime)= (-{C}/{K})[F_1(u^{\prime})+(1/2)({e_b}^2+{e_c}^2)F_2(u^{\prime})]$,
equation (\ref{w13}) can be approximately written as:

\begin{eqnarray}\label{w25}
\delta T(x^{\prime},y^{\prime})&\propto  & \int\exp[f(u^\prime)]\bigg [1 -\frac {C}{K}F_3(u^{\prime})\nonumber\\
&&    \frac{  {e_b}^2(x^\prime \cos\phi-y^\prime \sin\phi \cos\theta+z^\prime\sin\phi\sin\theta)^2 +{e_c}^2(y^\prime\sin\theta+z^\prime\cos\theta)^2}{2{r^\prime}^2}\bigg ]{\mathrm d}{z^\prime}\nonumber\\
& = & \int\exp[f(u^{\prime})]{\mathrm d}{z^\prime} -\frac{C}{K}\int\! \exp [f(u^{\prime})]F_3(u^{\prime})\nonumber\\
&& \bigg[ \frac{  {e_b}^2(x^\prime \cos\phi-y^\prime \sin\phi \cos\theta+z^\prime\sin\phi\sin\theta)^2 +{e_c}^2(y^\prime\sin\theta+z^\prime\cos\theta)^2}{2{r^{\prime}}^2}\bigg ]{\mathrm d}{z^\prime}\nonumber\\
& = & \int\! \exp [f(u^\prime)]{\mathrm d}{z^\prime}-\frac{C}{K{R_0}^2}[({e_b}^2\cos^2\phi){x^\prime}^2+({e_b}^2\sin^2\phi\cos^2\theta+{e_c}^2\sin^2\theta){y^\prime}^2 \nonumber\\
& & -(2{e_b}^2\sin\phi\cos\phi\cos\theta){x^\prime}{y^\prime}]\int\!\exp [f(u^\prime)]\frac{F_3(u^\prime)}{2{u^\prime}^2}{\mathrm d}{z^\prime}\nonumber\\
&&-\frac{C}{K{R_0}^2}({e_b}^2\sin^2\phi\sin^2\theta+{e_c}^2\cos^2\theta)\int\!
\exp[f(u^\prime)] {F_3(u^\prime)}   \frac{{z^\prime}^2}{2{u^\prime}^2}{\mathrm
d}{z^\prime}
\end{eqnarray}

The above approximation assumed that the non-circular part of
$\delta T$ is small due to the small values of $e_b$ and $e_c$.
The validity of this assumption depends on the value
of $C/K$, which is related to the concentration parameter of the
dark halo. In Figure 6, we showed the
comparisons between the approximate results [equation (A4) for isothermal
$\delta T$] and those from the
direct integrations of equations given in the previous section [equation (14)
for the isothermal $\delta T$] for the isothermal SZ (left)
and X-rays (right) for several different values of $C/K$.
It is seen that the two agree very well for reasonable $C/K$ values,
which are discussed
in Appendix B.

Let us define
\begin{eqnarray}\label{w26}
{\hat{F}}_{11}(\tilde u^\prime)  & = & \int\! \exp[f(u^\prime)]{\mathrm d}{z^\prime} \nonumber\\
                        & = & \int\! \exp {\bigg \{} -\frac{C}{K}\bigg [F_1(u^\prime)+\frac{{e_b}^2+{e_c}^2}{2}F_2(u^\prime)\bigg ] {\bigg \}}{\mathrm d}{z^\prime},
  \end{eqnarray}
\begin{eqnarray}\label{w27}
{\hat{F}}_{12}(\tilde u^\prime)  & = & \int\! \exp[f(u^\prime)]F_3(u^\prime)\frac{{z^\prime}^2}{2{u^\prime}^2}{\mathrm d}{z^\prime} \nonumber\\
                          & = & \int\! \exp {\bigg \{} -\frac{C}{K}\bigg [F_1(u^\prime)+\frac{{e_b}^2+{e_c}^2}{2}F_2(u^\prime)\bigg ] {\bigg \}}F_3(u^\prime)\frac{{z^\prime}^2}{2{u^\prime}^2}   {\mathrm d}{z^\prime},
\end{eqnarray}
\begin{eqnarray}\label{w28}
{\hat{F}}_{13}(\tilde u^\prime)  & = &\frac{C}{K} \int\! \exp[f(u^\prime)]\frac{F_3(u^\prime)}{2{u^\prime}^2}{\mathrm d}{z^\prime} \nonumber\\
                          & = &\frac{C}{K} \int\! \exp {\bigg \{} -\frac{C}{K}\bigg [F_1(u^\prime)+\frac{{e_b}^2+{e_c}^2}{2}F_2(u^\prime)\bigg ] {\bigg \}}\frac{F_3(u^\prime)}{2{u^\prime}^2}   {\mathrm d}{z^\prime},
\end{eqnarray}
and
\begin{equation}\label{w29}
C_{11} = -{e_b}^2\cos^2\phi,
\end{equation}
\begin{equation}\label{w30}
C_{12} = -({e_b}^2\sin^2\phi\cos^2\theta+{e_c}^2\sin^2\theta),
\end{equation}
\begin{equation}\label{w31}
C_{13} ={e_b}^2\sin{2\phi}\cos\theta.
\end{equation}
Then we have
\begin{eqnarray}\label{w32}
\delta T(x^\prime,y^\prime) & \propto & {\hat{F}}_{11}(\tilde u^\prime)-\frac{C}{K{R_0}^2}({e_b}^2\sin^2\theta\cos^2\phi+{e_c}^2\cos^2\theta){\hat{F}}_{12}(\tilde u^\prime)\nonumber\\
&&+\frac{1}{{R_0}^2}(C_{11}{x^\prime}^2+C_{12}{y^\prime}^2+C_{13}{x^\prime}{y^\prime}){\hat{F}}_{13}(\tilde u^\prime) \nonumber\\
& = & {\hat{F}}_{11}(\tilde u^\prime)-\frac{C}{K{R_0}^2}({e_b}^2\sin^2\theta\cos^2\phi+{e_c}^2\cos^2\theta){\hat{F}}_{12}(\tilde u^\prime) \nonumber\\
& &
+(C_{11}\cos^2{\theta^\prime}+C_{12}\sin^2{\theta^\prime}+C_{13}\cos{\theta^\prime}\sin{\theta^\prime}){\hat{F}}_{13}(\tilde u^\prime){\tilde u^{\prime 2}}
\end{eqnarray}

Comparing with equation (A3), we obtain
\begin{eqnarray}\label {w34}
\delta {
T}(\xi)\propto {\hat{F}}_{11}(\xi)-\frac{C}{K{R_0}^2}({e_b}^2\sin^2\theta\cos^2\phi+{e_c}^2\cos^2\theta){\hat{F}}_{12}(\xi).
\end{eqnarray}
Then to the linear order of $e_b^2$ and $e_c^2$, $\Delta$, $\Delta^\prime$ and $\Psi$ are determined by the following
relations
\begin{eqnarray} \label{w35}
\left\{
\begin{array}{l l}
\frac{ {\displaystyle \partial}{\displaystyle  {\hat F}_{11}}(  {\displaystyle {\tilde u^\prime}})}{\displaystyle {\partial {\tilde u^\prime}}}(-\frac{\displaystyle {\Delta} \displaystyle{\cos}^2{\displaystyle \Psi}+\displaystyle{\Delta ^\prime}\displaystyle{\sin}^2\displaystyle{\Psi}}{\displaystyle 2})=C_{11}{\hat F}_{13}{\tilde u^\prime}\\
\frac{ {\displaystyle \partial}{\displaystyle  {\hat F}_{11}}(  {\displaystyle {\tilde u^\prime}})}{\displaystyle {\partial {\tilde u^\prime}}}(-\frac{\displaystyle {\Delta} \displaystyle{\sin}^2{\displaystyle \Psi}+\displaystyle{\Delta ^\prime}\displaystyle{\cos}^2\displaystyle{\Psi}}{\displaystyle 2})=C_{12}{\hat F}_{13}{\tilde u^\prime}\\
\frac{ {\displaystyle \partial}{\displaystyle  {\hat F}_{11}}(  {\displaystyle {\tilde u^\prime}})}{\displaystyle {\partial {\tilde u^\prime}}}(-\frac{ \displaystyle {\Delta} \displaystyle{\sin}{\displaystyle 2\Psi}-\displaystyle{\Delta ^\prime}\displaystyle{\sin}\displaystyle{2\Psi}}{\displaystyle 2})=C_{13}{\hat F}_{13}{\tilde u^\prime}.\\
\end{array}
\right.
\end{eqnarray}

Explicitly, we have
\begin{equation}\label{w36}
\tan {2\Psi}=\frac{C_{13}}{C_{11}-C_{12}},
\end{equation}

\begin{equation}\label{w37}
\Delta=-[-C_{13}\sqrt{1+(\frac{C_{11}-C_{12}}{C_{13}})^2}+(C_{11}+C_{12})]M_1(\tilde
u^\prime),
\end{equation}

\begin{equation}\label{w38}
\Delta^\prime=-[C_{13}\sqrt{1+(\frac{C_{11}-C_{12}}{C_{13}})^2}+(C_{11}+C_{12})]M_1(\tilde
u^\prime),
\end{equation}
and
\begin{eqnarray} \label{w39}
{E}^2=1-\frac{1+\Delta}{1+{\Delta}^\prime}
=\frac{{\Delta}^\prime-\Delta}{1+{\Delta}^\prime}
\approx{\Delta}^\prime-\Delta=2C_{13}\sqrt{
 1+(\frac
 {C_{11}-C_{12}}{C_{13}})^2 } [-M_1(\tilde u^\prime)],
 \end{eqnarray}
where $M_1(\tilde u^\prime)=\tilde  u^\prime {\hat  F}_{13}(\tilde u^\prime)/{{\partial}_{\tilde u^\prime} {\hat F}_{11}(\tilde u^\prime)}$

%
%
%

For the other three cases, the only difference is showed up in
the $M$ factor, $M_i(\tilde u^\prime)=\tilde  u^\prime {\hat  F}_{i3}(\tilde u^\prime)/{{\partial}_{\tilde u^\prime} {\hat F}_{i1}(\tilde u^\prime)}$,
where $i$ indicates different observables.

For isothermal X-ray emission,
\begin{equation}\label{w47}
{\hat{F}}_{21}(\tilde u^{\prime})=\int\! \exp {\bigg \{}
-\frac{2C}{K}\bigg [F_1(u^{\prime})+\frac{{e_b}^2+{e_c}^2}{2}F_2(u^{\prime})
\bigg ] {\bigg \}}{\mathrm d}{z^\prime},
\end{equation}

\begin{equation}\label{w48}
{\hat{F}}_{23}(\tilde u^{\prime})=\frac{2C}{K} \int\! \exp {\bigg \{}
-\frac{2C}{K}\bigg [F_1(u^{\prime})+\frac{{e_b}^2+{e_c}^2}{2}F_2(u^{\prime})
\bigg ] {\bigg \}}\frac{F_3(u^{\prime})}{2u^{\prime 2}}   {\mathrm d}{z^\prime}.
\end{equation}

For polytropic SZ effect,
\begin{equation}\label{w49}
{\hat F}_{31}(\tilde u^{\prime})=
\int\!\bigg \{1-{C\over K_0}{\gamma-1\over \gamma}\bigg [F_1(u^{\prime})+\frac{{e_b}^2+{e_c}^2}{2}F_2(u^{\prime})-{\Phi_0\over C}\bigg ]\bigg \}^{\gamma/(\gamma-1)}{\mathrm d}z^\prime,
\end{equation}

\begin{equation}\label{w50}
{\hat F}_{33}(\tilde u^{\prime})={C\over K_0}\int\!\bigg \{1-{C\over K_0}{\gamma-1\over \gamma}\bigg [F_1(u^{\prime})+\frac{{e_b}^2+{e_c}^2}{2}F_2(u^{\prime})-{\Phi_0\over C}\bigg ]\bigg \}^{1/(\gamma-1)}\frac{F_3(u^{\prime})}{2u^{\prime 2}}{\mathrm d}z^\prime
\end{equation}

For polytropic X-ray emission,
\begin{equation}\label{w51}
{\hat F}_{41}(\tilde u^{\prime})=
\int\!\bigg \{1-\frac{C}{K_0}\frac{\gamma-1}{\gamma}\bigg [F_1(u^{\prime})+\frac{{e_b}^2+{e_c}^2}{2}F_2(u^{\prime})-{\Phi_0\over C}\bigg ]\bigg \}^{(3+\gamma)/[2(\gamma-1)]}{\mathrm d}z^\prime
\end{equation}

\begin{equation}\label{w52}
{\hat F}_{43}(\tilde u^{\prime})={C\over K_0}{3+\gamma\over 2\gamma}
\int\!\bigg \{1-\frac{C}{K_0}\frac{\gamma-1}{\gamma}\bigg [F_1(u^{\prime})+\frac{{e_b}^2+{e_c}^2}{2}F_2(u^{\prime})-{\Phi_0\over C}\bigg ]\bigg \}^{(5-\gamma)/[2(\gamma-1)]}\frac{F_3(u^{\prime})}{2u^{\prime 2}}{\mathrm d}z^\prime
\end{equation}

\section{Appendix}

The parameter ${C/ K}$ (${C/ K_0}$ for polytropic case)
is related to the characteristic
density of dark matter halos and the temperature information of ICM.
Specifically, we have
\begin{equation}\label{w50}
{C\over K}={4\pi G\delta_c \rho_{crit}R_0^2\over (k_BT_g/\mu m_p)}.
\end{equation}
The value of $C/K$ depends on boundary conditions.

Since we only consider nonspherical mass distribution up to
$e_b^2$ and $e_c^2$, $C/K$ can be estimated under the spherical
assumption. For isothermal ICM, the equilibrium equation tells us
\begin{equation}\label{w50}
\frac{dp_g}{dr}=-\rho_g\frac{d\Phi}{dr}\nonumber,
\end{equation}
and from it, we obtain
\begin{equation}\label{w50}
{C\over K}=-{u(d\ln{\rho_g})/(d\ln
u)\over {\int_0^u du^\prime} {u^\prime}^2/ [{u^\prime}^\alpha(1+u^\prime)^{3-\alpha}]},
\end{equation}
where $u\equiv r/R_0$. We adopt the boundary
condition that at the virial radius, the slope of ICM
traces that of the dark matter distribution. Then
\begin{equation}
\frac{C}{K}={c_{vir}[1+(2c_{vir})/(1+c_{vir})]\over \ln(1+c_{vir})-
c_{vir}/(1+c_{vir})},
\end{equation}
where $c_{vir}=r_{vir}/R_0$.

In the case of polytropic ICM,  following Komatsu \& Seljak (2001) we have
\begin{equation}
\frac{C}{K_0}=\frac{3}{\eta(0)}\frac{c_{vir}}{m(c_{vir})},
\end{equation}
where
$$m(c_{vir})=\ln(1+c_{vir})-\frac{c_{vir}}{1+c_{vir}},$$
$$s_\star=-\bigg [\alpha+(3-\alpha)\frac{c_{vir}}{1+c_{vir}}\bigg ],$$
and
$$\eta(0)=\gamma^{-1}\bigg \{-\frac{3}{s_\star}+3(\gamma-1)\frac{c_{vir}}{m(c_{vir})}\bigg [1-\frac{\ln(1+c_{vir})}{c_{vir}}\bigg ]\bigg \}.$$


\begin{figure}
\plotone{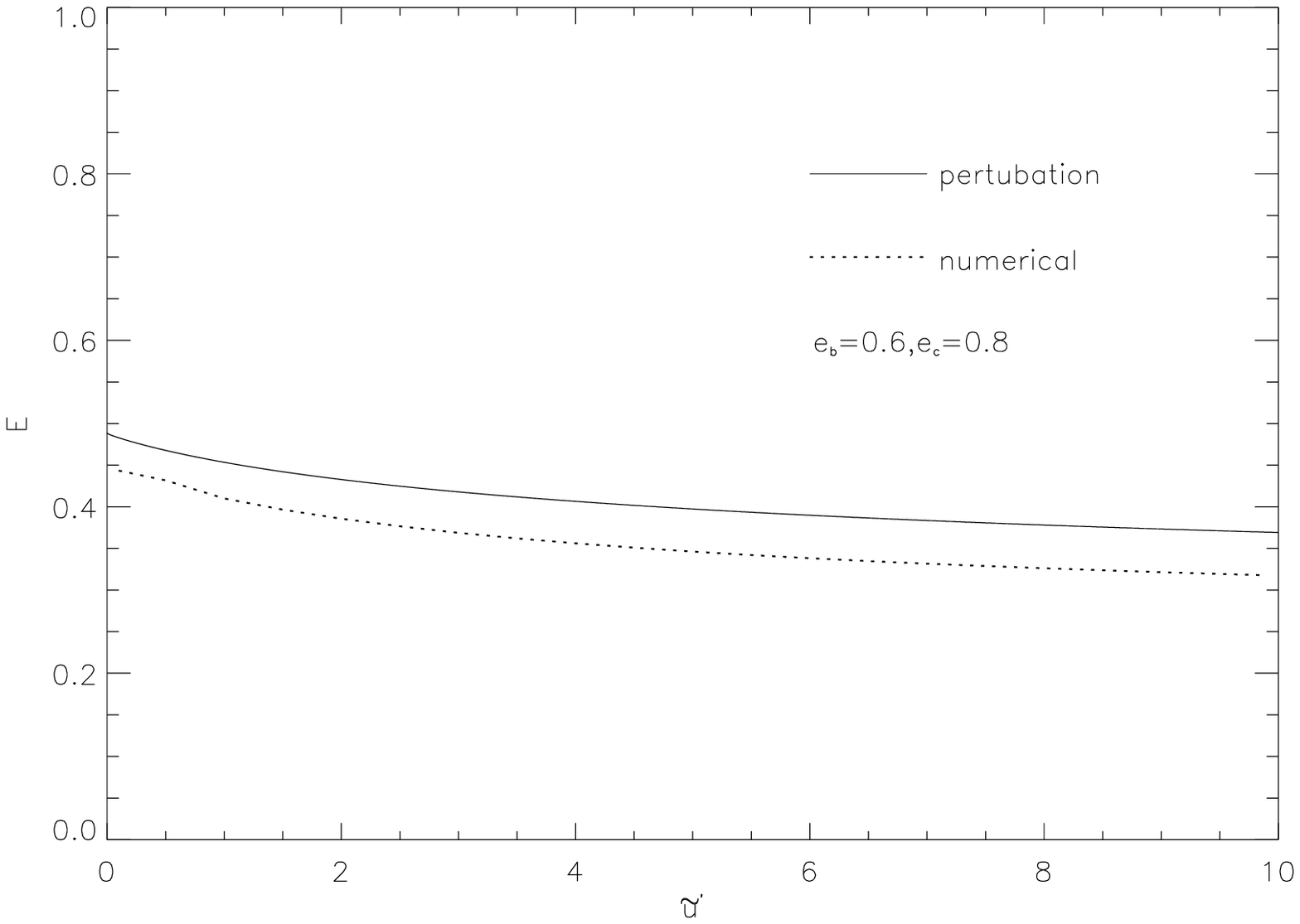}
 \caption{The eccentricity of isothermal SZ effect along the major axis.
 The cluster mass is $M=10^{14}h^{-1}M_\odot$, and $e_b=0.6$ and $e_c=0.8$.
 The line-of-sight direction is $\theta=0$ and $\phi=0$. The solid line
 is the perturbative results of equation (25), and the dotted line is
 the results derived from numerical calculations without
 small $e_b$ and $e_c$ approximation.
 \label{yg4}}
 \end{figure}

\begin{figure}
\plotone{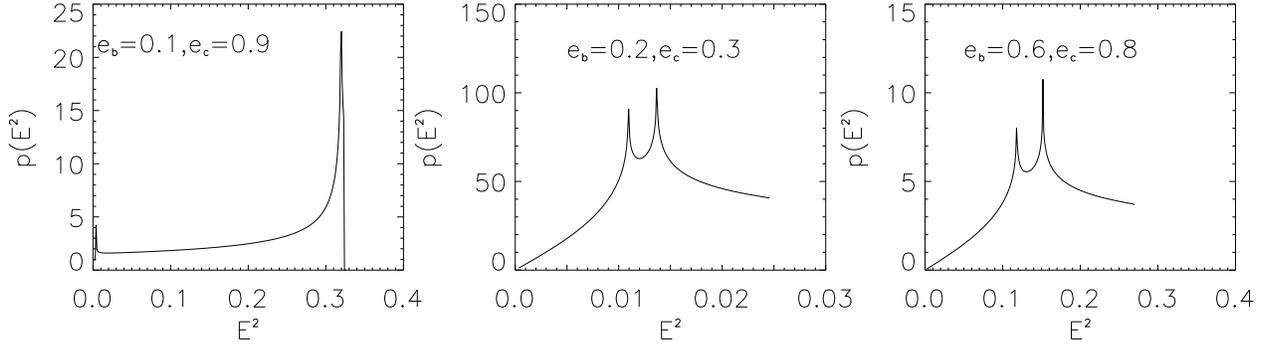}
\caption{The conditional probability $p(E^2|c_e,c/b,c/a)$
for isothermal SZ effects.
The mass of the cluster is $M=10^{14}h^{-1}M_\odot$, and the redshift
is $z=0$. The eccentricity is measured at $r_{vir}$.
Three sets of $(e_b,e_c)$ are given. For each set,
$c_e$ is taken as the average value calculated from the fitting
formula of Jing and Suto (2002).
\label{yg6}}
\end{figure}

\begin{figure}
\plotone{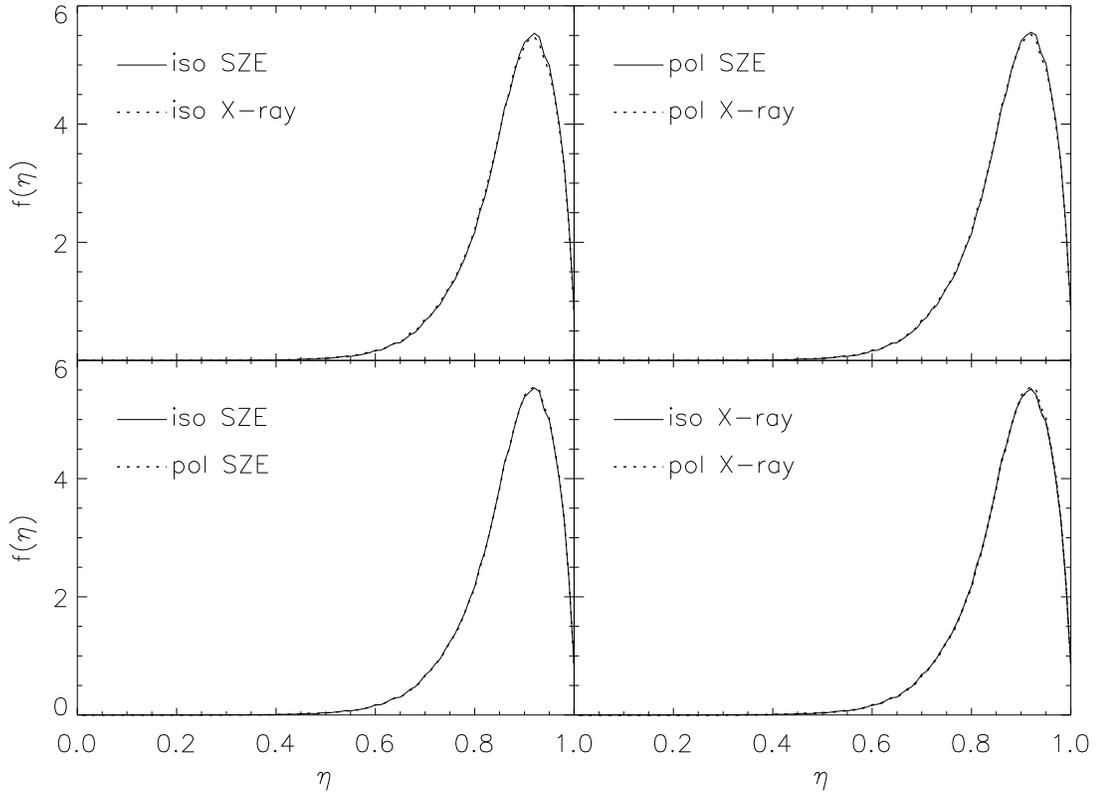}
 \caption{The probability function $f(\eta)$ of the axial ratio $\eta$
for $M=10^{14}h^{-1}M_\odot$ and $z=0$.
 \label{yg7}}
 \end{figure}

 \begin{figure}
 \plotone{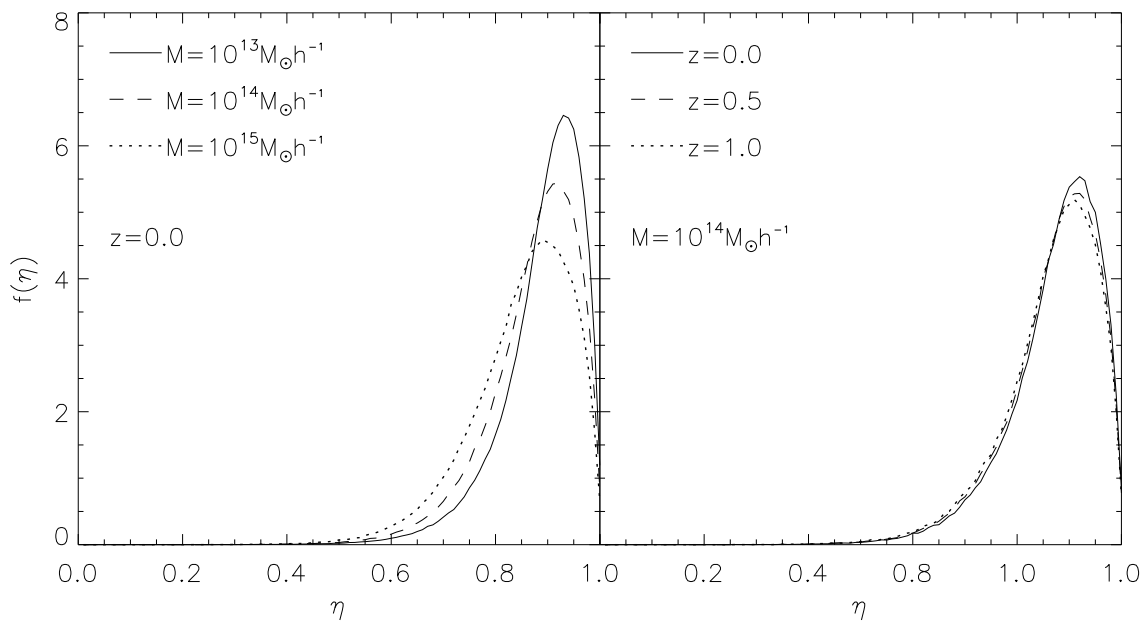}
  \caption{The probability function $f(\eta)$ for isothermal SZ effects.
  Left panel: The clusters are at $z=0$. The solid line is for
  $M=10^{13}h^{-1}M_{\odot}$. The dashed line is for $M=10^{14}h^{-1}M_{\odot}$.
  The dotted line is for $M=10^{15}h^{-1}M_{\odot}$.
  Right panel: $M=10^{14}h^{-1}M_{\odot}$. The solid, dashed, and dotted lines
  represent results of $z=0,0.5$ and $1.0$, respectively.
  \label{yg8} }
  \end{figure}

\begin{figure}
\plotone{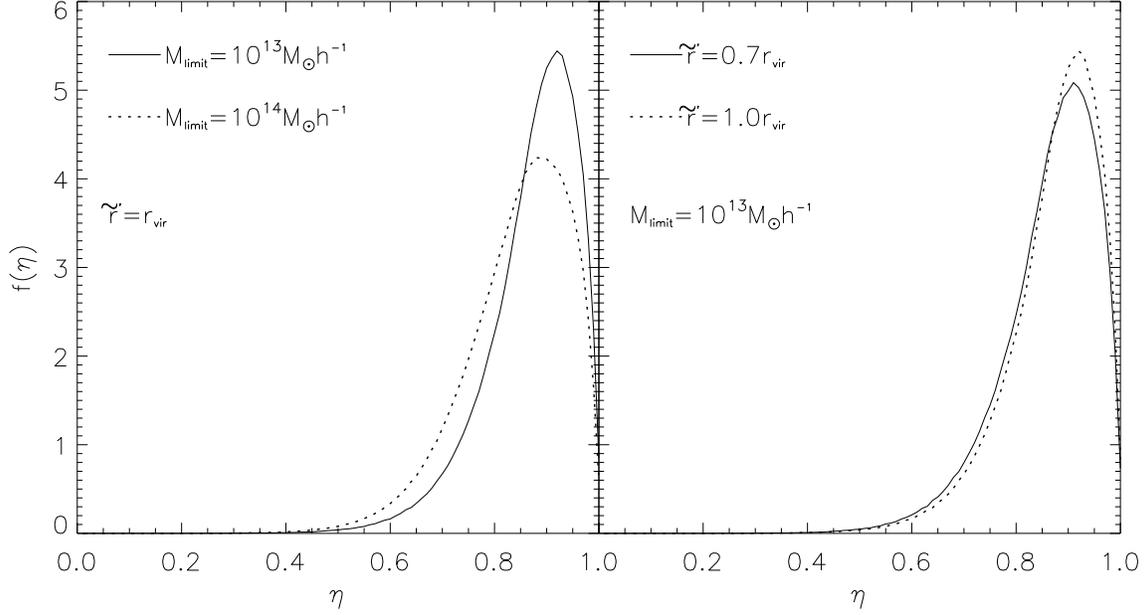}
\caption{The probability function $f(\eta)$ for mass-limited cluster samples.
Left panel: The solid line is for $M_{lim}= 10^{13}h^{-1}M_{\odot}$, and
the dotted line is for $M_{lim}= 10^{14}h^{-1}M_{\odot}$. The axial ratio
is measured at the virial radius $r_{vir}$. Right panel: The solid and dotted
lines are for the results that $\eta$ is measured at $0.7r_{vir}$ and $r_{vir}$,
respectively. The cluster mass $M=10^{13}h^{-1}M_\odot$.
\label{yg10}}
\end{figure}

\begin{figure}
\plotone{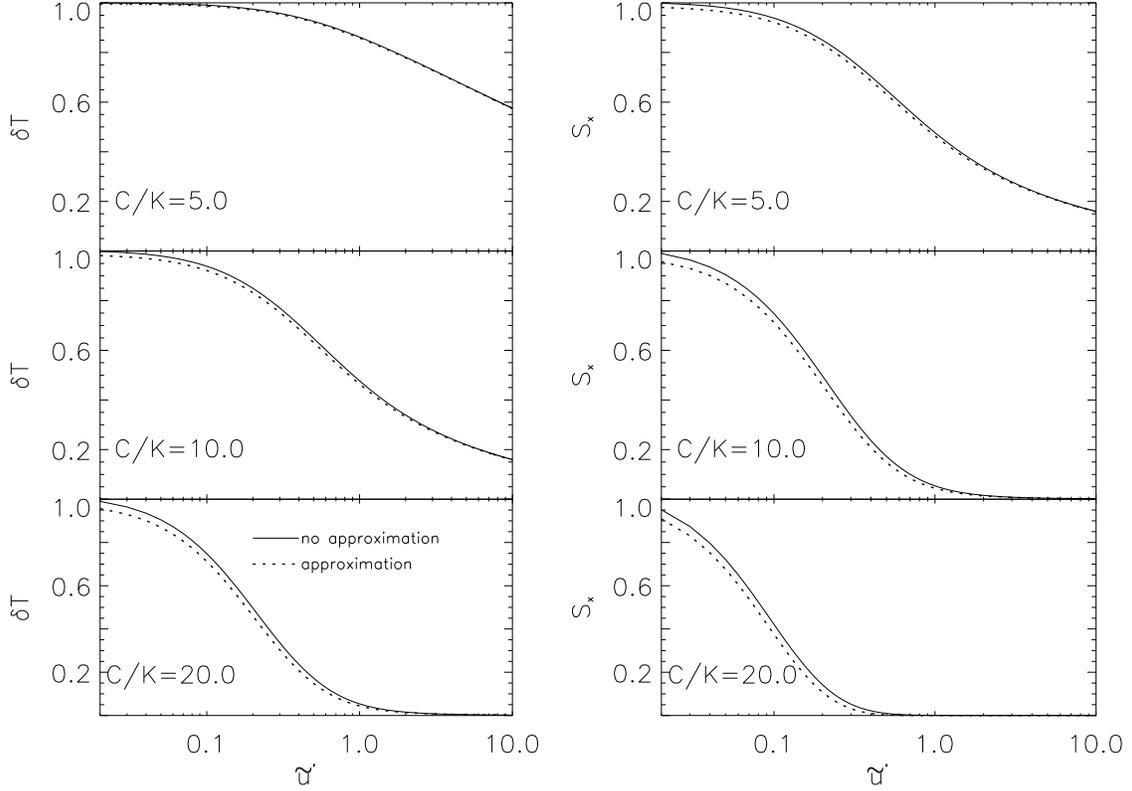} \caption{The profiles of SZ effects (left panels)
and X-ray emissions (right panels) along the major axis for isothermal ICM.
The cluster parameters and the line-of-sight parameters are the same
as those in Figure 1. The solid line in each panel represents the results
from direct integration along the line-of-sight [equation (14) for SZ effect].
The dotted line is calculated from the expansion [equation (25) for SZ effect].
The parameter $C/K=5,10$ and $20$, respectively.
\label{yg2}}
\end{figure}

\end{document}